\documentclass[letterpaper]{sig-alternate-2013}

\permission{\copyright 2017 International World Wide Web Conference Committee \\ (IW3C2), published under Creative Commons CC BY 4.0 License.}
\conferenceinfo{WWW'17 Companion,}{April 3--7, 2017, Perth, Australia.}
\copyrightetc{ACM \the\acmcopyr}
\crdata{978-1-4503-4914-7/17/04. \\
http://dx.doi.org/10.1145/3041021.3054176  \\
\includegraphics{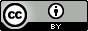}}

\usepackage{booktabs} 
\usepackage[T1]{fontenc}
\usepackage[utf8]{inputenc}
\usepackage{amsmath,amssymb}
\usepackage{cite}
\usepackage{subfigure}
\usepackage{nameref,hyperref}
\usepackage{enumerate}
\usepackage{color}
\usepackage{enumitem}
\usepackage{graphicx}
\usepackage{fancyvrb}
\usepackage{fancyhdr}
\usepackage{algorithm,algorithmic}
\usepackage{multirow}
\usepackage{balance}

\def\pprw{8.5in}
\def\pprh{11in}

\begin{document}

\title{
Achievement and Friends: Key Factors of Player Retention Vary Across Player Levels in Online Multiplayer Games
}

\numberofauthors{1}  
\author{
\alignauthor Kunwoo Park$^*$~~~~Meeyoung Cha$^{**}$~~~~Haewoon Kwak$^\dag$~~~~Kuan-Ta Chen$^\ddag$\\
\affaddr{$^*$Graduate School of Web Science Technology, School of Computing, KAIST, South Korea}\\
       \affaddr{$^{**}$Graduate School of Culture Technology, KAIST, South Korea}\\
       \affaddr{$^\dag$Qatar Computing Research Institute, Hamad Bin Khalifa University, Qatar}\\
       \affaddr{$^\ddag$Academia Sinica, Taiwan}\\
       \email{\{kw.park,meeyoungcha\}@kaist.ac.kr~~~haewoon@acm.org~~~ktchen@iis.sinica.edu.tw}
}

\maketitle
\begin{abstract}
Retaining players over an extended period of time is a long-standing challenge in game industry. Significant effort has been paid to understanding what motivates players enjoy games. While individuals may have varying reasons to play or abandon a game at different stages within the game, previous studies have looked at the retention problem from a snapshot view. This study, by analyzing in-game logs of 51,104 distinct individuals in an online multiplayer game, uniquely offers a multifaceted view of the retention problem over the players' virtual life phases. We find that key indicators of longevity change with the game level. Achievement features are important for players at the initial to the advanced phases, yet social features become the most predictive of longevity once players reach the highest level offered by the game. These findings have theoretical and practical implications for designing online games that are adaptive to meeting the players' needs.
\end{abstract}

\keywords{Player retention; virtual life trajectory; player level; online multiplayer games; longevity}

\section{Introduction}
 
Player retention is a critical and long-running quest in online game industry. What makes players stay happy in a game and follow through its scenario? What makes them continue the game even after having reached the highest level offered? To answer these questions, researchers have studied the  motivations of game players for over a decade~\cite{yee2006motivations,williams2008plays,debeauvais2011if}. Studies based on theoretical investigations, user surveys, and log data analyses have identified several factors that are critical to retention. For example, players are known to find enjoyment in games from completing missions, empowering through growth and level ups, forming communities, competing against other players, discovering plots and characters, and more.

Previous studies have tried to group these motivating factors and measure their relative strengths in retaining players. Researchers have found that players can be grouped into a few set of clusters based on their game motivations as action-social (i.e., players who enjoy fast-paced scenario with player interaction), mastery-achievement (i.e., players who indicate interests in narrative, expression, and world exploration), and immersion-creativity (i.e., players who appeal to strategic game plays, taking on challenges, and becoming powerful). Game designers carefully implement  reward mechanisms of each motivation type throughout game scenarios to meet the needs of different players. Existing work has assumed that the relationship between players and motivations is rigid (e.g., does not change over time) and is irrespective of the players' virtual life phases. 

This study brings a multifaceted aspect to this important question by examining retention over various phases of individual lifetime. We assume that one's potential and capacity to enjoy a game changes over time, and hence the need and the ability to achieve higher levels quickly and to socialize within games for cooperative shifts must be different for each individual. By observing in-game behavior logs throughout various phases of real game players, this paper sets out to answer the following research questions:
\begin{itemize}
\item For each phase within an online multiplayer game, what are the characteristics of players who achieve the next higher levels and get retained?
\item Why do some individuals continue to play even after having reached the max level?
\end{itemize}

We utilize logs gathered from one of the oldest massively multiplayer online role-playing games (MMORPGs) in the world, Fairyland Online in Taiwan. We gained access to the complete set of actions of 51,104 individuals, describing their achievement logs (quests and level ups), financial logs (gaining wealth), as well as social logs (chats among players). Myriads of action logs on tens of thousands of individuals who ultimately achieved different levels and played the game for different amounts of time allow us to design a natural experiment on the lifetime retention problem. We identify the factors attributing to game longevity through the detailed log analysis and make the following observations:

\begin{enumerate}
\item Achievement features are important for players during the initial to advanced phases; players who are achievement-oriented and gather large amounts of rare items and virtual money are more likely to be retained and succeed in achieving the next levels. 

\item Achievement-related traits, however, are no longer as important for players who reach the max level. Social features become the most predictive of success and longevity beyond this point.

\item Having strong social relationships (measured by the number of friends) is a good indicator of player retention and their effect continues to show significance through virtual life phases of players.
\end{enumerate}

Our findings bring theoretical and practical implications for studying and designing online games. The finding that a player's needs vary over one's virtual life trajectory needs to be carefully addressed by further research and game designs. In particular, findings on longevity of the max level players is new. This finding is particularly important as their behaviors have not been studied much, even though expert players are valuable to the user ecosystem. 

\section{Related Works}

Since Bartle~\cite{bartle1996hearts} defined the four-type player taxonomy based on motivation in text-based games, there have been numerous efforts to understand why people join and continue to play online games. Among them is Yee's findings on three motivational components---achievement, social, and  immersion---based on factor analysis of survey results on Bartle's player types. This study also identified that motivations can vary across different demographics. While general MMO players are found to be achievement-oriented~\cite{williams2008plays}, females were more likely to play online games to have social relationship with other players. The reason that players' motivation has been studied for decades is partly because of its ultimate connection to player retention. 

Among recent findings, Debeauvais et al.~\cite{debeauvais2011if} asked World of Warcraft players about their motivation for play and game usage patterns through questionnaire surveys and found that socially-motivated players are more likely to discontinue games while achievement-oriented players tend to continue. Borbora et al.~\cite{borbora2011churn} built a prediction model of player  motivation from log data. From data mining experiments using player activity logs from Everquest II, they found achievement is a dominant motivation for predicting player churn (i.e., opposite of player retention). Above studies consistently report that achievement is a major motivation for retention in online games. On the other hand, some studies found social activity to be more important for retention. Based on the log data of EverQuest II, a study showed social influences from peers help predict player retention better~\cite{kawale2009churn}. 

Most recently, another group of researchers observed that game interactions such as interacting with toxic players can have negative impacts on retention in League of Legends~\cite{shores2014identification}.
As cyberbullying has been considered as one of the factors that make players annoyed, feel fatigued, and even leave the game~\cite{mulligan2003developing}, there have been much efforts to define, detect, and prevent toxic playing in online games~\cite{kwak2015exploring,blackburn2014stfu,Kwak2015linguistic}.  However, in this work, we do not investigate the effect of cyberbullying on player engagement due to the limitation of our dataset.

While many studies have put efforts to contribute to understanding player retention, much of the findings have been drawn from a snapshot view---players aggregated by demographic features yet not considering \textit{how they grow over time} within a game. Like human life itself, players face different challenges and engage in specific actions depending on their levels. For example, Ducheneaut et al.~\cite{ducheneaut2006alone} observed that online game players are more likely to play alone in an early stage, but become socially active at higher level. Players need to collaborate with one another to defeat strong monsters or complete difficult quests as their level elevates. Moreover, players enjoy an entirely different in-game experiences once they achieve the maximum level, as they become socially active without consuming much game content~\cite{ducheneaut2006building}. This means that factors leading to higher levels or being retained may be different across the entire player lifetime within online games. 

However, little attention has been paid to characteristics of churners over different phases of players. To the best of our knowledge, only one study by Shores et al.~\cite{shores2014identification} investigated how indicators of player retention compare between new-joiners and experts in a MOBA (multiplayer online battle arena) game, and there is room for improvement. First, churn types can be examined for more than two groups. Player behaviors continuously change with level, and hence it is more natural to observe the whole picture of player life trajectories. Second, comprehensive data provide richer views. While the study relied on add-ons to gather data, the kinds of data that could be gathered externally is limited. Utilizing in-game logs provide full picture of player behaviors that might be important for predicting retention. Third, the studied game is a specific type that does not capture the growth of players naturally. MOBA game is repeated matches of the ground which have importance on team formations~\cite{kim2016proficiency}, yet MMORPG allows characters to explore and grow as individuals.

\begin{figure}[t]
{
	\centering
	\includegraphics[width=0.97\linewidth]{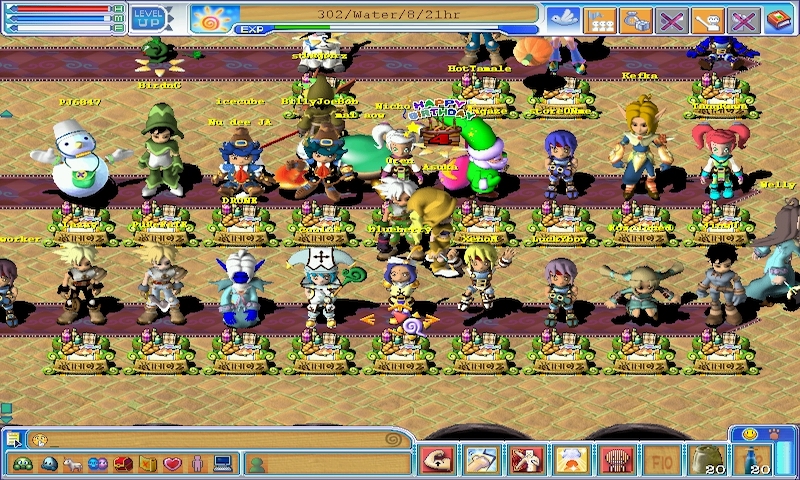}
	\vspace{-1mm}
	\caption{A screenshot of Fairyland Online~\cite{fairylandfig}}
	\label{fig:fairyland}
}
\end{figure}

\section{Dataset}

Fairyland Online is one of the longest serviced MMORPGs, which has been played in Taiwan and other nearby countries since its launch in 2003. As depicted in Figure~\ref{fig:fairyland}, the game is set on a virtual world that sets on fairy tales. Players can create their own avatars by choosing a race among human, elf, and dwarf and a gender of either female or male. On the virtual world, players explore their kingdoms, complete quests by fighting with monsters, and form social relationships with one another. Every action in the game is recorded in the game servers with accurate timestamps. Thanks to the Larger Network Technologies that serviced Fairyland Online, we gained access to the log data describing all actions that have been performed in the game. 

{
\begin{figure*}[t]
\hspace*{-5mm}
\centering
\minipage{0.45\textwidth}%
  \includegraphics[width=0.97\linewidth]{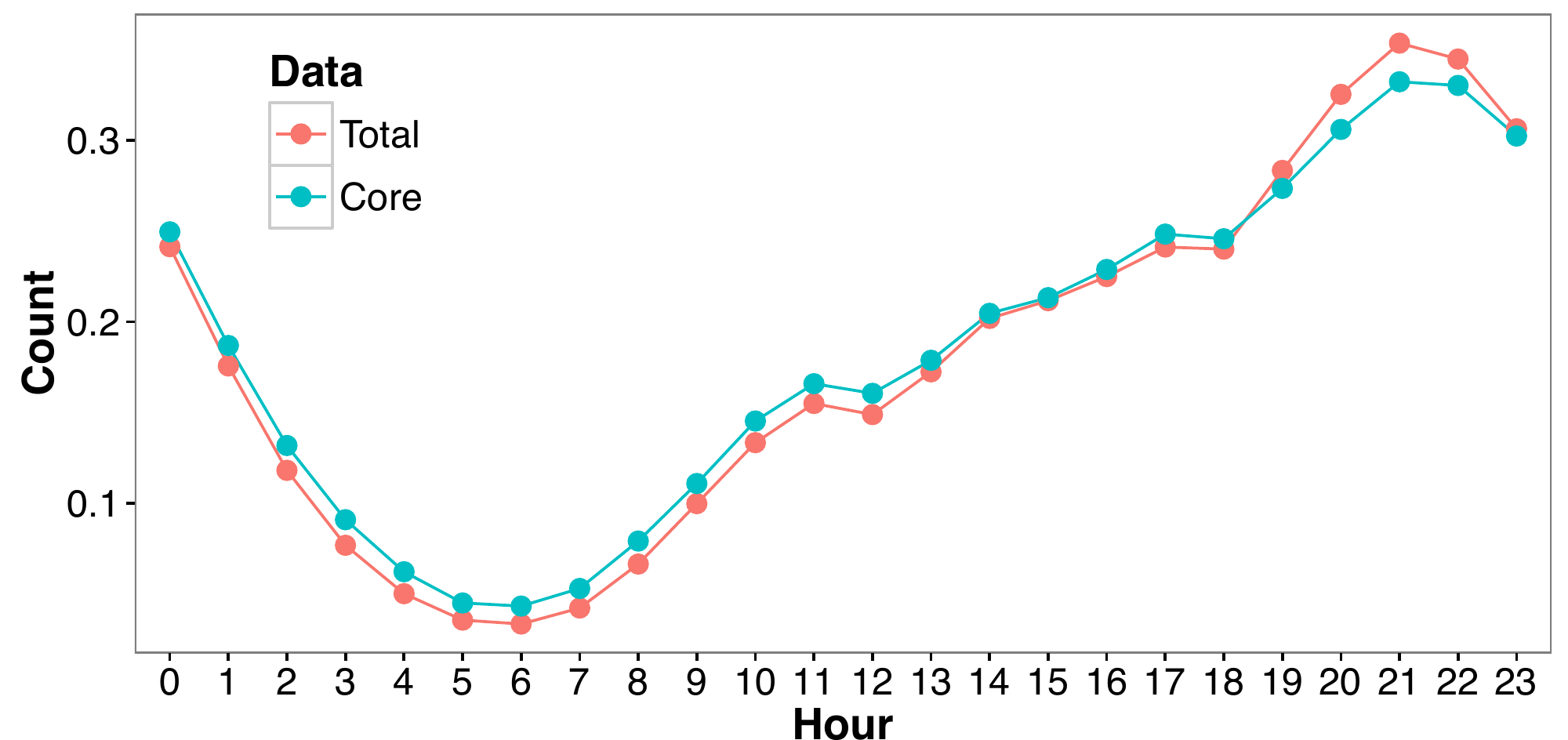}
\centering (a) Game access within a day
\endminipage
\minipage{0.45\textwidth}
  \includegraphics[width=0.97\linewidth]{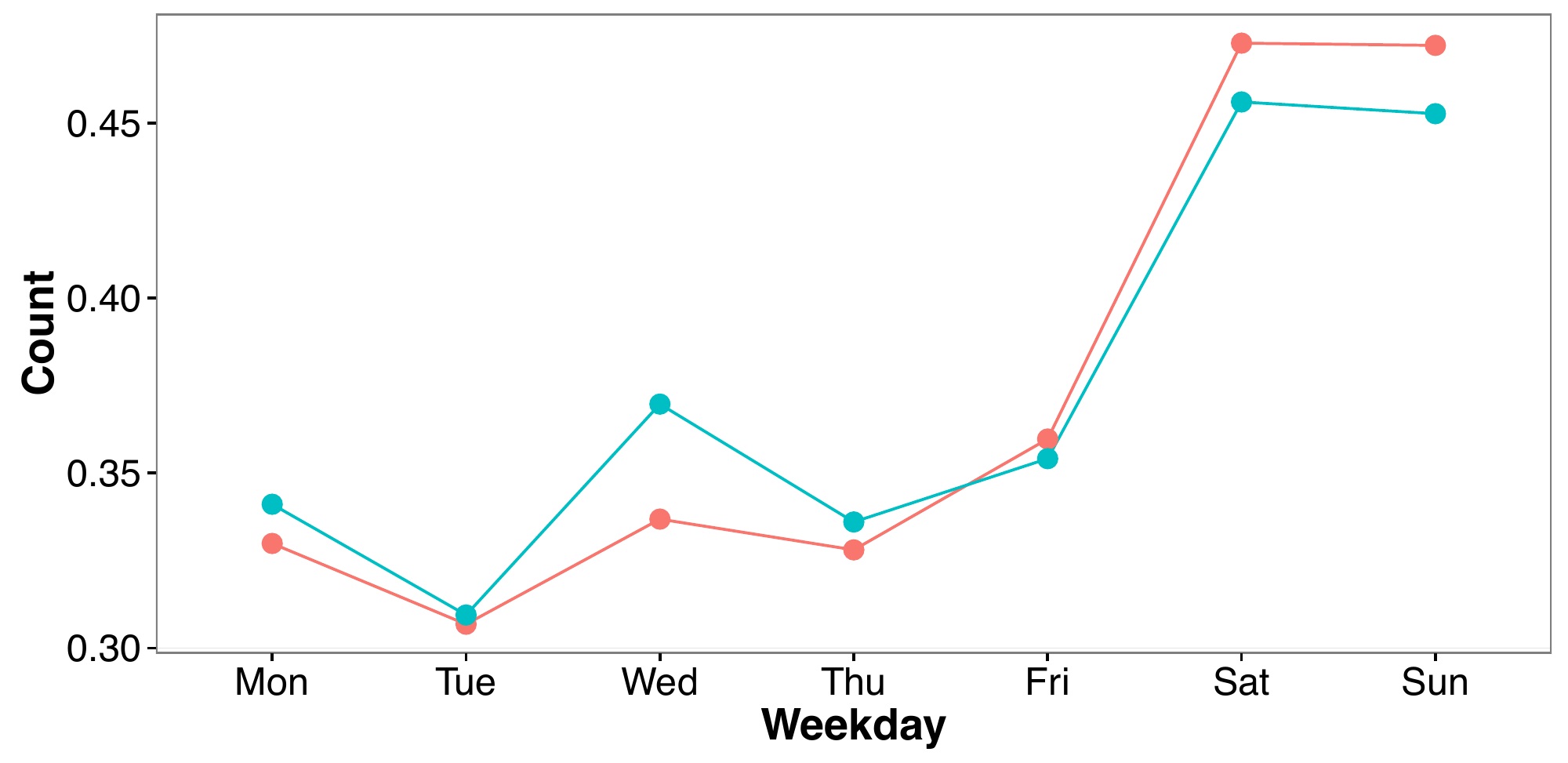}
\centering (b) Game access within a week
\endminipage
\caption{Temporal aspects from the normalized log counts of the entire and \textsf{Core} datasets}\label{fig:total_counts}
\end{figure*}
}

Fairyland Online servers log three different types of datasets. Firstly, there are logs related to achievement experience points (e.g., learn skills, completing quests). When a player gains enough experience points, his or her level will increase.  Secondly, there are a group of actions related to gaining or losing wealth on the virtual world (e.g., buy or sell items, earn or use money). Game items can be retrieved by defeating monsters or by purchasing with virtual money. The last logs are about chats among players. There are four different channels to chat: Say, Whisper, Family, and Party. Say is a public channel through which a player can communicate with multiple companions. Messages on Say channel are broadcasted. Thus, if someone writes a message through Say channel, whoever in a virtual proximity can see it. Whisper is a private channel between two players. Because Whisper channel is private, no one except for the speakers and receivers can overhear those messages. Family is a dedicated channel for players who belong to the same `family', which is equivalent to what is called  `guild' in other MMORPGs~\cite{steinkuehler2006everybody}. Party is the mode of communication for short-term groups. For privacy concerns, the chat contents themselves were encrypted.

The three kinds of datasets we received covered different time spans. We consolidated them to find a common overlapping period, over which we gain a full view of the achievements, financial, and social activities within the game. The overlapped portion covered nearly 60 million activity instances logged for 51,104 game players. We refer to this final complete dataset as \textsf{Core} in this paper and describe its statistics in Table~\ref{table:dataset}. The table also shows the unique number of game players and record instances logged over the three original datasets. 

{
\begin{table}[h] 
  \centering 
  \caption{Dataset summary}
\begin{tabular}{ccccc}
\toprule
  Type & Period & Players & Records\\\midrule
  \textsf{Core} & 2003 4 21--2003 7 8 & 51,104 & 59,466,664\\
  \textsf{All} & 2003 2 21--2004 5 31 & 157,812 & 262,711,811\\
   \bottomrule
\end{tabular}
\label{table:dataset}
\end{table}
}

Given the timestamps of actions, we may infer when  players accessed the game throughout a day and a week. The 24-hour plot shown in Figure~\ref{fig:total_counts}(a) depicts that, for both the entire logs and for the \textsf{Core} period, players show strong diurnal patterns. The game was played more during late night with a peak between 8 pm and 10 pm. A disproportionately fewer players logged on during early morning times. In the morning, the highest time is around 11 am (before lunch time), then an increasing number of players join in the afternoon and evening hours. Differences between the entire log and \textsf{Core} is marginal, indicating that the final \textsf{Core} data set we study is representative of the entire log in terms of temporal patterns. 

Figure~\ref{fig:total_counts}(b) presents the normalized daily access pattern across the data, again for the entire log and for \textsf{Core}. We find the game is played 1.4--1.5 times more frequently on weekends than during the weekdays. Later, we will investigate such detailed temporal features of players (e.g., weekend vs weekday oriented, most active time of a day) in predicting player retention. Note that the temporal patterns seen here have also appeared in other game studies~\cite{pittman2007measurement}, which suggests that the studied Fairyland Online shares commonalities with other representative MMORPGs.

\section{Methodology}

\subsection{Phase Definition}

The key objective of this research is to know what keeps a player in a game over various phases (and in particular toward the very end) of his or her virtual life. To answer this question, we start with arbitrary grouping of one's lifetime. In this work, we do not consider the very first phases of a game (i.e., beginners), which is a specific subset of our problem. We focus on players who have spent enough time to be accustomed to the rules of the game and determine what factors positively or negatively contributed to reaching the next level. 

Players in Fairyland Online can have a level between the lowest of 1 and the highest of 50. Among such players, we decide target users to represent each phase of the online game by one quantity: the observable level ranges for each player during \textsf{Core} (i.e., level of 10--15, 20--25, 30--35, 40--45, and 45--50). Players belonging to the 10--15 level group must have achieved a level of at least 15 and have their traces since level 10 visible in the \textsf{Core} period.

Table~\ref{table:phase} describes the five representative phases of virtual life that we examine in this paper. Where the exact division among groups lies is less of importance in this work. Rather, we are more interested in finding trends that become more prominent among the advanced, long-term game players. For each phase, we define success differently. The first four phases of 1--4 in Table~\ref{table:phase} allow us to examine whether each player \textit{successfully achieves} the next five levels. For phase 1 (i.e., level 10--15), we consider players who ultimately reach level 20 as success and otherwise as unsuccessful. For phase 5, the success is defined as whether the game is able to \textit{keep a given player} or not. We decided a player is churned when he or she becomes inactive for consecutive 90 days over the next 270 days. To test the validity of the length of days to decide user churn, we varied the number of days from 30 to 180 and they showed similar results with Table~\ref{table:result_phase5}, which will be presented in the next section.

{
\begin{table}[h]
  \centering
  \caption{Phase information along with grouping criteria, definition of success at a given phase, the number of corresponding individuals, and the probability of success within the group}
\begin{tabular}{ccccc}
\toprule
  Phase & Levels & Success defined & Num & Prob\\\midrule
  1 & 10--15 & Achieve level 20 & 3,818 & 0.6612\\
  2 & 20--25 & Achieve level 30 & 1,739 & 0.7483\\
  3 & 30--35 & Achieve level 40 & 1,370 & 0.8299\\
  4 & 40--45 & Achieve level 50 & 674 & 0.5638\\
  5 & 45--50 & Retained after level 50 & 221 & 0.4198\\
  \bottomrule
\end{tabular}

\label{table:phase}
\end{table}
}

Out of the 51,104 individuals in the \textsf{Core} dataset, we only consider players whose observable levels are within the given ranges as described by Phase 1--5 as target players. We also ensure that each player has at least 31 days observable within in \textsf{Core} after the end of the observation level. For instance, for Phase 1, we ensure that individuals had at least 31 days after they reached level 15 in the log. This gives ample time for them to meet the success criteria (i.e., achieving level 20 for Phase 1). The buffer time of 31 days was determined from the log analysis. We investigated how long it takes to achieve the higher level for each phase listed in the table for a subset of 91 players who joined and achieved the maximum level during the \textsf{Core} data period. 
These players took 13.92 days on average with the standard deviation of 1.29 and a maximum of 30.97 days to achieve 5 level ups (i.e., from level 45 to level 50). Thus, we set the buffer length as 31 days. For Phase 5 players, however, we did not enforce this buffer length, as they no longer need to achieve a higher level. For these players, churn was measured over the entire log period (beyond the \textsf{Core} period). From Phase 1 to Phase 5, we identified 3818, 1739, 1370, 674, and 221 players meeting the above criteria, respectively. Note that a player can belong to multiple groups so long as he or she meets the criteria.

The last column of Table~\ref{table:phase} displays the probability of success for each phase, where the success criterion is also listed in the table itself. The fraction of players who success in a given phase is the highest among Phase 3 players (i.e., individuals who were in level 30--35) and is the lowest for Phase 5 players (i.e., individuals who reach the highest level offered). Nonetheless, the success probability or the retention rate is considerably stable, remaining over 0.4 throughout the five phases.

\subsection{Studied Features}

We utilize a total of 16 features for predicting user retention. The features are divided into three major categories based on their characteristics: temporal, achievement-related, and social. 

\subsubsection{Temporal Category}

The temporal features describe when individuals played the game. Temporal patterns not only reveal how often a user plays the game (e.g., every day vs. once a week) but also reveal certain demographic traits. For example, play time can be used to infer which players are likely students (e.g., peak immediately after the school hours) or which players likely work regular hours (e.g., playtime starts only after the typical business hours).  
We extract two features like below:

\begin{itemize}
\item Frequent hours (\textit{morning}, \textit{working\_hour}, \textit{evening}, and \textit{night\_owl}): To capture playing patterns, we measured how frequently a player accesses the game with the following time blocks. We define 4 variables that represent specific playing patterns: morning from 6 am to 9 am, working hours from 9 am to 6 pm, evening from 6 pm to midnight, and night owl from midnight to 6 am. We conducted vector normalization on those variables to remove the effects of the number of engaged days.

\item Weekday vs weekends (\textit{weekends}): the fraction of playtime contributed from weekends. 

\end{itemize}

\subsubsection{Achievement-Related Category}

Many studies have reported the importance of achievement as a goal in player retention~\cite{borbora2011churn,debeauvais2011if}. To measure its effect, we utilize the following features related to in-game achievements.

\begin{itemize}
\item Possessed item count (\textit{item}): the total count of owned items measured by subtracting the number of item-losing logs from item-gaining logs. This quantity is a proxy of in-game achievements.  
\item Rare items on hand (\textit{rare\_item}): Owning rare items can be more important achievement than general item counts. To decide which items are rare, we approximated chances of getting an item by counting the frequency over the whole item frequencies measured from \textsf{Core}. Then, we considered items appeared with the probability lower than 0.01 to be rare items. We measured the number of rare items on hand in the same manner as we did to count for items on hand.
\item Amount of money on hand (\textit{money}): the amount of virtual money that each player has, calculated by the differences between money-gaining logs and money-losing logs. 
\item Level of difficulty (\textit{difficulty}): the level of difficulty, measured by a combination of the number of deaths and  broken items. The appropriate level of difficulty has been considered as an important element for user engagement in online games~\cite{chanel2008boredom}.
\item Performance (\textit{performance}): The performance of achieving level can represent level of motivations on achievement. We measured it by changing the sign of time length of observation period, because each user can take different time to achieve 5 levels of the observation period based on his or her performance. A larger performance value hence indicates that a player leveled up quickly.
\end{itemize}

\subsubsection{Social Category}

Social features are another important group of indicators for player engagement~\cite{kawale2009churn,ducheneaut2006building,tyack2016appeal}. Below we describe the list of social features we tested in this paper.

\begin{itemize}
\item Number of social interactions (\textit{num\_social}): the number of all messages that a player sent through any channel---a measure of social activeness.
\item Response rate (\textit{response\_rate}): the probability of giving responses when a player received messages from an unknown player---a measure of social openness. 
\item Number of friends (\textit{friends}): To figure out effects of social interactions in a more detail, we define friendships. Based on whisper logs, we counted the number of distinct days paired communications take place. If a player has paired communications with another player for at least three different days, we considered the communication partner to be a friend. To measure this variable, we counted the number of friends who communicated at the observation period as a feature for retention.
\item Number of non-friends (\textit{nonfriends}): We considered those who have paired communications but are not friends to be non-friends. The number of non-friends were observed from the observation period of each phase.
\item Friends' level (\textit{friends\_level}): To represent the level of friends, we got the median level of friends who have communicated on the observation period. 
\item Non-friends' level (\textit{nonfriends\_level}): the median level of non-friends, who have communicated with the player during the observation period.
\item Number of max-level friends (\textit{friends\_maxlevel}): the number of friends who communicated with the player and already achieved the maximum level at the moment of communication on the observation period.
\item Number of max-level non-friends (\textit{nonfriends\_maxlevel}): the number of non-friends who have paired communication with the player and already achieved the maximum level when the communication happens.
\item Is a member of a family (\textit{has\_family}): a binary variable whether the player belongs to a family, which is a membership-based group. We inferred it based on whether a user has sent messages through Family channel.
\end{itemize}

\subsection{Player Retention Model}

A logistic regression model was used to determine factors that affect player longevity. The regression model helps us investigate how various indicators attribute to player retention across different life phases within the virtual world, while allowing us to control for the effects of other variables. Hence we choose to use interpretable models in this paper rather than implementing other kinds of prediction models that might achieve higher performance. 

Prior to analyses, a step was taken to balance the data. Because the success rate at each phase is biased toward one side, we employed an over-sampling technique to prepare an equal number of success and fail cases for each phase. All variables were scaled to have a mean of 0 and a standard deviation of 1. In addition, since variables of regression models can turn out to be significant simply due to a large number of predictors, we performed variable selections using Lasso~\cite{tibshirani1996regression} by choosing the lambda whose cross-validated mean squared error is within one standard error of the minimum. In the results section, we report findings of the regression fitting after this feature selection step.

\section{Results}

This research assumes the important indicators of the player retention vary throughout the different phases in Fairyland Online. To test this idea, for each phase of the game, we fit the logistic regression model of the successful cases and unsuccessful cases (as defined in Table~\ref{table:phase}) across the 16 features from three categories (i.e., temporal, achievement, and social). We compare the relative importance of each category in predicting player retention. 

\subsection{Low- to Medium-level Patterns}

Among the five phases of the game level, here we focus on Phase 1 to Phase 3, which are logs of players who have  become accustomed to the game. Our goal is to understand what kinds of players are more likely to be retained and further succeed in achieving the next levels. The three phases were observed among more than a thousand individuals. Below we only list the final set of features deemed as meaningful (out of the 16 features) for each of the three phases, after the Lasso variable selection step. 

{
\begin{table}[h] \small
  \centering
  \caption{Results for Phase 1 players (level 10--15)}
\begin{tabular}{llrr}
  Category & Predictor & Estimate & Significance\\\toprule
  & (Intercept) & 0.062 &  \\\midrule
  \multirow{2}{*}{Temporal} & \textit{night\_owl} & 0.424 & ***\\
   & \textit{weekends} & -0.244 & *\\\midrule
  Achievement & \textit{performance} & 0.371 & ***\\\midrule
  \multirow{3}{*}{Social} & \textit{friends} & 0.539 & ***\\
   & \textit{nonfriends} & -0.331 & ***\\
   & \textit{nonfriends\_maxlevel} & -0.09 & **\\
  \midrule
Model $\chi^2$ & & 435.8 & *** \\\bottomrule
\multicolumn{4}{r}{*:p$<$0.05, **:p$<$0.01, ***:p$<$0.001} \\

\end{tabular}
\label{table:result_phase1}
\end{table}
}

Table~\ref{table:result_phase1} presents the fitted results for Phase 1, which shows the estimates and significance of variables. The table also shows the model $\chi^2$ value based on the likelihood ratio test with a null model. We see that at least one feature from the temporal, achievement, and social categories appears as significant. Among the temporal features, \textit{night\_owl} is positively associated with the success, while \textit{weekends} is negatively correlated. This means individuals who mainly played after midnight and during weekdays (but not just on weekends) were more likely to reach the next levels---This could indicate that at an early stage, time dedication is an important marker of success. Among the achievement features, \textit{performance} (i.e., negative quantity of the playtime) increases the probability of success in that players with speedy game style are more likely to be retained and achieve the next levels. As many studies found, achievement is one of the main motivations for continuing to play online games~\cite{borbora2011churn,debeauvais2011if}. Our analysis also confirms that in an early virtual stage achievement help players reach the next levels without leaving games. 

Among the social features, \textit{friends} is positively associated with success, while \textit{nonfriends} is negatively associated with success. This may indicate that players with many friends yet fewer weak social ties are more likely to achieve the next levels. This trend supports findings from several studies on the importance of social interactions in games~\cite{kawale2009churn,ducheneaut2006building}. In contrast, the negative effect of \textit{nonfriends} is interesting. It may indicate that communication with too many random users may be harmful for long-term engagement. Furthermore, \textit{nonfriends\_maxlevel} shows negative association with the success rate in that players who communicate with many max-level non-friends are less likely to continue with the game. Communicating with too advanced non-friends may be a negative experience on future engagement, because players can feel left behind~\cite{tandoc2015facebook}.

{
\begin{table}[h] \small
  \centering
  \caption{Results for Phase 2 players (level 20--25)}
\begin{tabular}{llrr }
  Category & Predictor & Estimate & Significance\\\toprule
   & (intercept) & -0.016 & \\\midrule
  \multirow{3}{*}{Temporal} & \textit{morning} & -0.096 & * \\
   & \textit{evening} & 0.072 & \\
   & \textit{night\_owl} & 0.184 & ***\\
   \midrule
  \multirow{3}{*}{Achievement} & \textit{item} & -0.203 & **\\
   & \textit{money} & 0.252 & ***\\
   & \textit{performance} & 0.497 & ***\\
  \midrule
  \multirow{6}{*}{Social} & \textit{friends} & 0.53 & ***\\
   & \textit{nonfriends} & -0.302 & ***\\
   & \textit{friends\_level} & -0.095 & *\\
   & \textit{friends\_maxlevel} & -0.177 & ***\\
   & \textit{nonfriends\_maxlevel} & 0.124 & *\\
   & \textit{has\_family} & -0.129 & **\\
  \midrule
Model $\chi^2$ & & 281.34 & *** \\\bottomrule

\end{tabular}
\label{table:result_phase2}
\end{table}
}

Phase 2 players indicated several more number of significant variables related to retention, as shown in Table~\ref{table:result_phase2}. Among the temporal features, \textit{night\_owl} is positively associated with success, while \textit{morning} is not. It seems still important to devote extra time on the game after midnight for achieving higher levels, while playing the game since early morning (i.e., 6--9 am) seem a not effective strategy for further engagement. Among the achievement features, \textit{performance} is again positively associated. We newly found \textit{money} to be positively correlated, yet \textit{item} is negatively associated. Accumulating in-game money increases the probability of the continued usage at Phase 2, because it may become difficult to quit the game after one gathers a large sum of virtual money. In addition, virtual wealth means one's ability to upgrade game avatars, which helps achieve the next levels easier. Those two possible explanations can be linked to the success of achieving more levels. However, simply owning many items decrease the chance of success. 

Among social features, \textit{friends} is again positively associated with the success, while \textit{nonfriends} is not. This finding implies that having more friends and fewer weak social relationship is linked to helping players achieve higher levels. Also, \textit{has\_family} is newly found to be significant with a negative estimate at this stage. In other words, players are less likely to succeed when they joined a family. This trend also supports the importance of focusing on strong social relationship to the continued usage of online games. Lastly, we observed that \textit{friends\_maxlevel}, and \textit{friends\_level} are negatively associated with success. This trend can be similarly explained with the negative association of \textit{nonfriends\_maxlevel} for prediction of Phase 1.

{
\begin{table}[h] \small
  \centering 
  \caption{Results for Phase 3 players (level 30--35)}
\begin{tabular}{llrr}
  Category & Predictor & Estimate & Significance\\\toprule
    & (intercept) & -0.017 & \\\midrule
  \multirow{3}{*}{Temporal} & \textit{working\_hour} & -0.082 & \\
  & \textit{night\_owl} & 0.251 & ***\\
   & \textit{weekends} & -0.199 & ***\\\midrule
  \multirow{5}{*}{Achievement} & \textit{item} & -0.274 & **\\
   & \textit{rare\_item} & 0.292 & ***\\
   & \textit{money} & 0.265 & **\\
   & \textit{difficulty} & 0.125 & **\\
   & \textit{performance} & 0.686 & ***\\\midrule
  \multirow{8}{*}{Social} & \textit{num\_social} & -0.234 & **\\
   & \textit{response\_rate} & -0.106 & \\
   & \textit{friends} & 0.607 & ***\\
   & \textit{nonfriends} & -0.28 & ***\\   
   & \textit{friends\_level} & -0.133 & **\\
   & \textit{nonfriends\_level} & 0.114 & *\\
   & \textit{nonfriends\_maxlevel} & -0.077 & \\
  \midrule
Model $\chi^2$ & & 375.67  & *** \\\bottomrule

\end{tabular}
\label{table:result_phase3}
\end{table}
}

Players in Phase 3 exhibit similar trends as in Phase 2 (Table~\ref{table:result_phase3}). Among temporal features, \textit{night\_owl} is positively correlated, yet  \textit{weekends} is not---This pattern is similar to what we have seen before in earlier phases. Among the achievement features, \textit{performance} is again important for predicting player retention. There are some new trends; while \textit{item} is still negatively associated, players who gather larger amounts of rare items (i.e., \textit{rare\_item}) are likely to succeed. With the positive association of \textit{money}, this finding supports the claim that owning virtual wealth is related to the success of achieving higher level. In addition, \textit{difficulty} (i.e., the number of deaths and broken items) was found to be a positive indicator for success. Once the game reaches a certain stage, an appropriate level of difficulty may help players better enjoy games, as reported in a previous study~\cite{chanel2008boredom}. 

Next, from social features, \textit{friends} is positively correlated, while  \textit{nonfriends} and \textit{num\_social} are not. Again this finding demonstrates the importance of having communication with close friends rather than simply being socially active. Lastly, among variables on whom users talked to, \textit{nonfriends\_level} was newly found as a positive estimator. We hypothesize that players who ask for more help to other players of higher level are more likely to succeed. As reported in previous works~\cite{nardi2007learning,fields2009connective}, communicating with experts can sometimes be helpful in online games because they share knowledge, useful tactics, and strategies that are critical in proceeding with the next phases. Because this process does not require having any strong relationships with those with high level, \textit{nonfriends\_level} is a positive indicator yet \textit{friends\_level} may remain to have the opposite effect. As discussed in the results seen in earlier phases, having social relationships with users with max level or higher may give detrimental effects on future engagement. 

In summary, we found two consistent trends from the regression analysis of low- to medium-level phases (i.e., Phase 1-3). One is that performance on achieving levels and playing patterns related to devoting more times increase the probability of success for achieving more levels. These findings can be connected to the importance of motivation on achievement for player retention. Another is that players who have more friends yet fewer weak social relationships were more likely to be engaged in Fairyland Online continuously. Playing games with friends may give positive effects on achieving more levels. These findings are consistent with previous findings on player retention on other games~\cite{borbora2011churn,debeauvais2011if,kawale2009churn,ducheneaut2006building}.

\subsection{High-level Patterns}

Players who reached a level 40 or above out of 50 in Fairyland Online may be considered advanced users. What are the factors that lead to successfully reaching the endgame for these advanced players? Table~\ref{table:result_phase4} displays the regression result for Phase 4 players. At this very last stage, the only meaningful feature left after the Lasso feature selection is  \textit{performance} (i.e., -1$\times$play time). Players who enjoy speedy game and are quick at leveling up are more likely to succeed to reach the max level. It is interesting to see that achievement-related feature alone is a critical factor of success. 

{
\begin{table}[h] 
  \centering
  \caption{Results for Phase 4 players (level 40--45)}
\begin{tabular}{llrr}
  Category & Predictor & Estimate & Significance\\\toprule
   & (intercept) & -0.007 & \\\midrule
  Achievement & \textit{performance} & 0.485 & ***\\\midrule
Model $\chi^2$ & & 40.505 & *** \\\bottomrule

\end{tabular}
\label{table:result_phase4}
\end{table}
}

Once players reach the max level, a different story unfolds. In contrast to the Phase 4 players, the only meaningful feature of longevity left at this stage is the social category, where \textit{friends} is the only significant indicator for retention for players who reach the highest level. This finding suggests that having a substantial number of friends is consistently important in determining who will continue to play online games even after completing all missions. Note that this variable was also important during the earlier phases, further suggesting the importance of social interactions for player retention from earlier stages to the endgame. As found in previous studies~\cite{ducheneaut2006building}, online games become more of a social space after the max level. To be engaged in such online games in the long run, players must have constructed strong social relationships from early on in their virtual lives.

{
\begin{table}[h] 
  \centering 
  \caption{Results for Phase 5 players (level 45--50)}
\begin{tabular}{llrr}
  Category & Predictor & Estimate & Significance\\\toprule
   & (intercept) & 0.035 & \\\midrule
  Social & \textit{friends} & 0.966 & ***\\
  \midrule
Model $\chi^2$ & & 44.172 & *** \\\bottomrule

\end{tabular}
\label{table:result_phase5}
\end{table}
}

\subsection{Trajectory Over Lifetime}

Having examined the factors of player retention step-wise, we now jointly view trends over the entire life stages in the Fairyland Online game. The set of features examined are from three main categories: temporal, achievement-related, and social. Which of these categories are important for predicting player retention at each phase? To answer this question, we compared the relative importance of the three categories in predicting player retention via training separately on each categorical features. For testing, 5-fold cross validation was used on the unbalanced original dataset with keeping the distribution of target labels. Then, we conducted over-sampling for each split to be balanced. We applied this sampling technique after each split to prevent for same instances to be both in training and test set for each split. We finally measured the area under the ROC curve (AUROC) of logistic regression classifiers using each set of features.

\begin{figure}[t]
{
	\centering
	\includegraphics[width=0.99\linewidth]{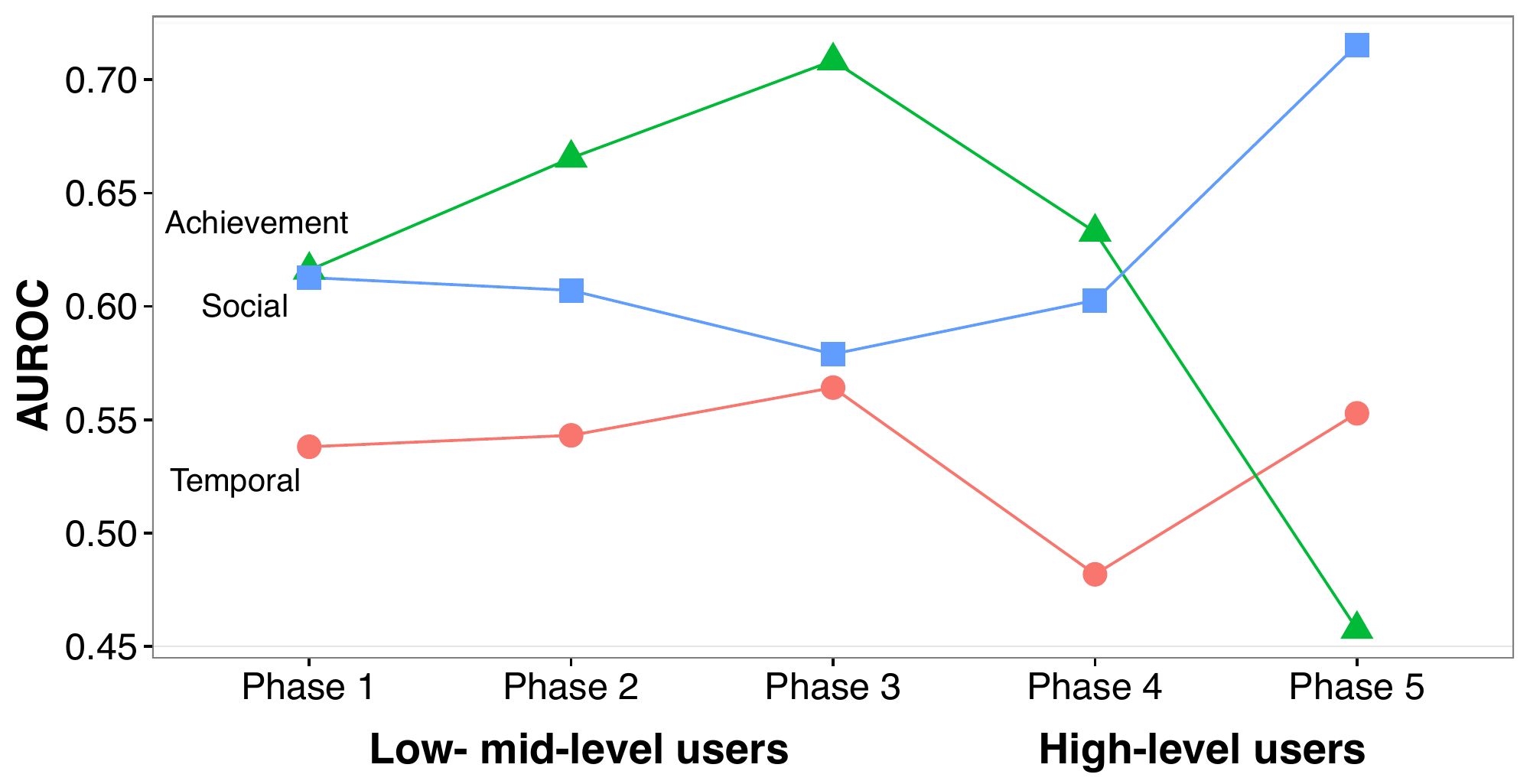}
	\vspace{-1mm}
	\caption{Prediction accuracy across the five phases seen by the area under the ROC curve (AUROC) of logistic regression for each feature category}
	\label{fig:auc_comp}
}
\end{figure}

Figure~\ref{fig:auc_comp} presents the changes of the AUROC values of the three categories over the five phases. The AUROC value is between 0 and 1, where a value of 1 means the prediction model is perfect. A prominent trend we see is the role of achievement features that show the best performance during the early to late phases of the game (i.e., Phase 1 to 4). The social category shows a comparable trend to the achievement category for Phase 1, 2, and 4. This category then becomes the most important in Phase 5 (i.e., the max level players), at which point the other features are no longer important. The temporal features are, for most of the phases, better than random guessing (i.e., AUROC of 0.5), although they do not show a big gain against the baseline. We discuss implications of these findings in the next section.

\section{Discussion \& Conclusion} 

Maintaining user base of a substantial size is critical for many companies in running their services. Companies across various industries (e.g., telecommunication companies~\cite{backiel2016predicting}, health app providers~\cite{park2016persistent}, and so on) have put their efforts to understand characteristics of people who discontinue services and predict them in advance based on data mining approaches. Game industry and researchers also noticed the importance of the player retention problem, and many studies hence tried to understand player motivations~\cite{yee2006motivations,williams2008plays}, behavioral characteristics~\cite{ducheneaut2006alone,ducheneaut2006building}, and to build prediction models based on studied  characteristics~\cite{borbora2011churn,kawale2009churn}. 

\begin{figure}[t]
{
	\centering
	\includegraphics[width=0.97\linewidth]{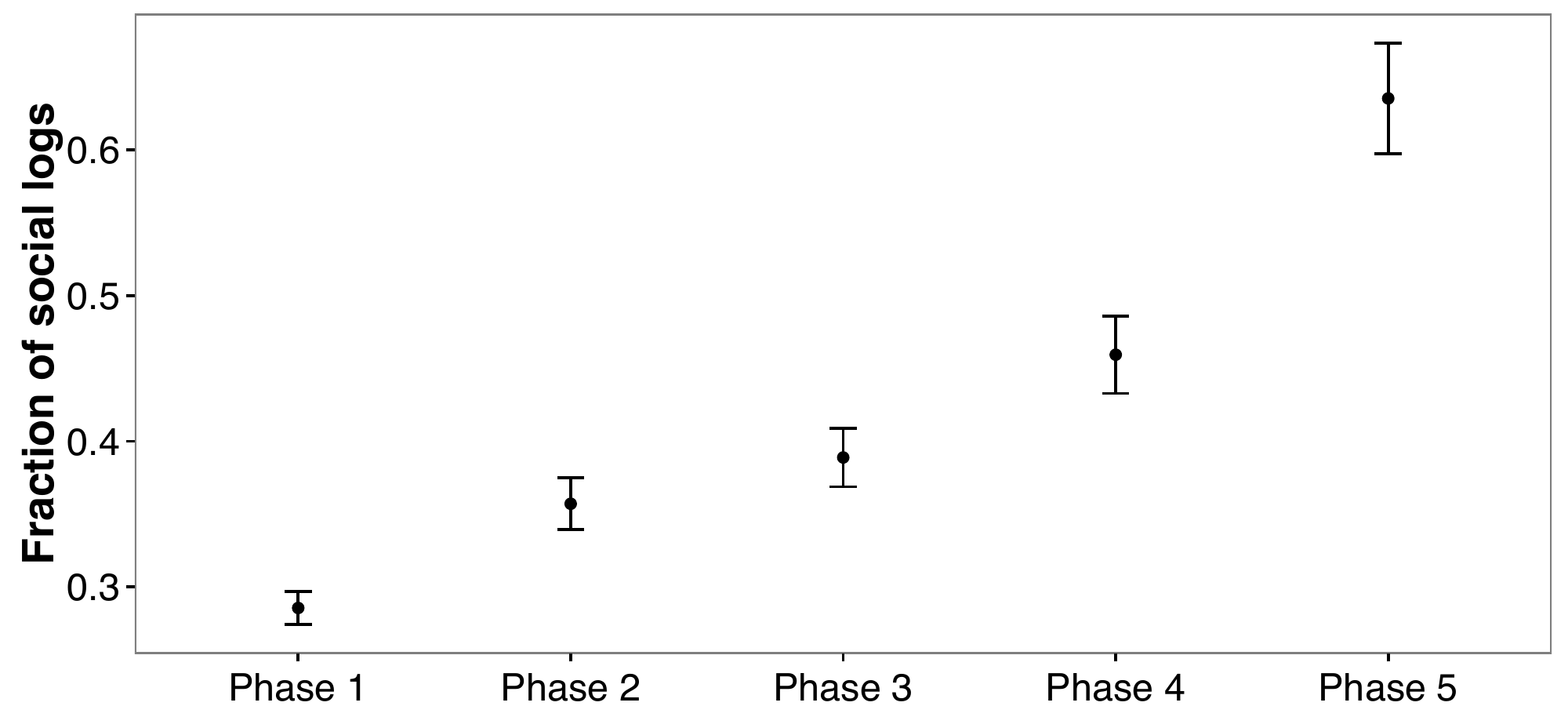}
	\vspace{-1mm}
	\caption{The fraction of logs related to social activities out of the entire logs at different life phases}
	\label{fig:social_frac}
}
\end{figure}

However, existing studies did not disentangle user groups and conducted analyses without considering player levels. Because game designs of MMORPGs let players to have a certain amount of activities as level increases~\cite{ducheneaut2006alone}, game players face different challenges as their level increases and this evolution can affect user retention. As in Figure~\ref{fig:social_frac}, we also observed that social interactions increase as level increases in our dataset. Thus, to precisely understand indicators for player retention, effect of features should be separately measured across different virtual life phases in online games. Another aspect that has received little attention is retaining individuals who have reached the highest level offered by the game. These expert players not only help newbies adapt to the game, but also are a major source of profit for game industry. Therefore, retaining the max-level players is a critical problem. 

Motivated by these missed opportunities, this research aimed to answer two research questions:  (i) what are the indicators for player retention over different phases of players and (ii) how does the relative importance of retention features change over the game phases. Through a series of quantitative analyses using 51,104 individuals based on in-game logs, we have made several key  findings for the question. Theses results are important for the following reasons. Firstly, we noted that the key indicators of longevity change with player phases. This finding implies that other studies on user behavior also need to consider phases of gamers. Secondly, our findings have practical implications to online game developers, as they need to carefully consider the changing needs of players over various life stages. Game designers may offer fast achievement-oriented scenarios at the beginning, while motivate players to form strong social relationships long before they reach any advanced level. We note that these suggestions are hypothetical, because observations indicate correlation not causality. Future studies can conduct controlled experiments or qualitative studies to further test causal relationship of feature effects. Another implication is that game industry may apply these findings to construct churn prediction models. For example, prediction models could be constructed separately for each phase and hence better capture signals of churning individuals.

On top of the above findings, we also found significant indicators observed for certain phases. Playing after midnight was positively associated with the success of Phase 1 to 3, while playing in the early morning is negatively associated with the continued usage. There may exist certain playing patterns, which can be linked to player retention. Also, obtaining rare items or larger amount of money increased the chance of success in low- to medium-level. Owning virtual wealth may be helpful to achieving more levels, or it could make them feel commitment to keep playing online games by having a large amount of wealth in the virtual world. Lastly, we found significant indicators on whom a player talked to. For example, the level of friends was a significant indicator during the initial phases. This finding implies that social network positively affects retention when individuals form interactions with partners of appropriate levels. Because these findings are newly found in this study, predicting player retention can be better improved with further investigation on those variables.

This paper has several limitations. Among them is the use of a single data source. Every MMORPG has different game elements and player traits, and hence our findings can not be directly generalized to other online games. Nonetheless, we expect Fairyland Online is representative of a typical MMORPG in terms of the temporal trends, which shows similarity to other  games~\cite{pittman2007measurement}. In the future we hope to replicate the study with other online game logs. Another limitation is that, although we tried extensive features across three different categories based on the related literature, there can exist missing features which might be critically linked to player retention. For example, the number of churned friends was found to be an indicator of player churn in one online game~\cite{kawale2009churn}. Due to limited data, we could not employ this feature for analysis. In a future work, we hope to look into a longer time period and investigate the effects of other possible indicators including churned friends. Lastly, we did not investigate players who are at their very early stages (i.e., level 1-10). This was because the initial level up in Fairyland Online was fairly easy and there was not much data associated with this time period. However, new-joiners are of great interest to game industry because they are critical to increasing user base and future studies can delve deeper into the behaviors of new-joiners across different games. 

\section*{Acknowledgement}

Cha and Park were supported by the Ministry of Trade, Industry \& Energy (MOTIE, Korea) under Industrial Technology Innovation Program (No.10073144), `Developing machine intelligence based conversation system that detects situations and responds to human emotions'.

{
\balance

}

\end{document}